\documentclass[preprint2]{aastex}
\pdfoutput=1
\usepackage{graphics}
\usepackage{natbib}
\usepackage{graphicx}
\usepackage{amssymb}
\usepackage{amsmath}
\usepackage{epstopdf}
\usepackage{float}
\usepackage{url}

\newcommand{\ltsimeq}{\raisebox{-0.6ex}{$\,\stackrel
        {\raisebox{-.2ex}{$\textstyle <$}}{\sim}\,$}}
\usepackage{url}
\usepackage{epstopdf}

\begin{document}

\title{Seeing Double at Neptune's South Pole}

\shortauthors{S. H. Luszcz-Cook et al.}

\author{
  S. H. Luszcz-Cook\altaffilmark{1},
  I. de Pater\altaffilmark{1},
  M. \'Ad\'amkovics\altaffilmark{1},
  and H. B. Hammel \altaffilmark{2}}

\altaffiltext{1}{Astronomy Department, University of California, 
Berkeley, CA, USA}

\altaffiltext{2}{Space Science Institute, Boulder, CO 80303, USA}


\begin{abstract}
Keck near-infrared images of Neptune from UT 26 July 2007 show that the cloud feature typically observed within a few degrees of Neptune's south pole had split into a pair of bright spots. A careful determination of disk center places the cloud centers at $-89.07 \pm 0 .06 ^{\circ} $ and $-87.84  \pm 0.06 ^{\circ}$ planetocentric latitude. If modeled as optically thick, perfectly reflecting layers, we find the pair of features to be constrained to the troposphere, at pressures greater than 0.4 bar. By UT 28 July 2007, images with comparable resolution reveal only a single feature near the south pole.  The changing morphology of these circumpolar clouds suggests they may form in a region of strong convection surrounding a Neptunian south polar vortex.

\vspace{\baselineskip}
\vspace{\baselineskip}
\vspace{\baselineskip}
\end{abstract}

%
%

\section{Introduction}

The atmosphere of Neptune is active. In addition to global changes on decadal scales \citep{lock06,hammel07a}, individual clouds evolve on time scales as short as hours \cite[e.g.][]{limaye91,srom95,srom01}. Near-infrared (NIR) images, which probe Neptune's lower stratosphere and upper troposphere in reflected sunlight, reveal bright bands and cloud features against a dark disk. The pattern of zonal circulation, which is most pronounced at mid latitudes, continues to high southern latitudes, where a bright, unresolved cloud feature is present within a few degrees of the pole. This feature has been seen since the Voyager 2 era \citep{smith89}.

Interest in Neptune's global dynamics has been reignited by mid-IR observations of enhanced emission over the south pole relative to the rest of the planet \citep{hammel07b,orton07}. This enhancement suggests a temperature increase of 4-5 K in the south polar region, and has led these authors to draw parallels with Saturn, where the summer pole is similarly warm in the stratosphere and troposphere \citep{orton05}. Recently it has been observed that both of Saturn's poles exhibit a localized hot spot surrounded by cooler zones, forming circulation cells that make up part of a planet-wide circulation \citep{fletcher08,dyudina08}. A global dynamical pattern has also been inferred for Neptune's atmosphere \citep{martin08}; it remains to be seen if this pattern includes a polar circulation cell like those on Saturn.

\begin {table*}[htb!]
\begin{center}
 \caption{Observations}
 \begin{minipage}{\textwidth}
 \vspace{\baselineskip}
 \small{
\begin{tabular}{l  l l l l l l l}
\hline
Date 		& Time (UTC)	& Filter 	&	Central $\lambda$ ($\mu$m)$^a$	&Bandpass ($\mu$m)$^a$ & Airmass		& T$_{exp}$ (sec) & N$_{exp}$\\
\hline
26 July 2007\\
\ \ Neptune	& 11:37              & J      	& 1.25           	& 0.163               	&  1.21"		&60			\ 	&3\\
\ \ Neptune	& 11:43             & Kp	   	& 2.12           	& 0.351                      	&  1.21"		&60				&3\\
 \ \ Neptune	& 11:49              & H     	& 1.63           	& 0.296                       &  1.21"		&60				&5\\
\ \ Neptune	& 11:58              & H      	& 1.63           	& 0.296                       &  1.22"		&60				&5\\
\ \ Neptune	& 12:07              & H      	& 1.63            	& 0.296                       &  1.22"		&60				&5\\
\ \ HD 22686	&   15:10            & J     	& 1.25   		& 0.163                       &  1.33"		&6				&3\\
\ \ HD 22686	&    15:13          & H      	& 1.63         	& 0.296                       &  1.32"		&6				&3\\
\ \ HD22686	&      15:15         & Kp      	& 2.12          	& 0.351                       &  1.31"		&10				&3\\
\ \ SAO 146732	&     15:21          & Kp     	& 2.12           	& 0.351 		    	&  1.22"		&5				&3\\
\ \ SAO 146732	&       15:23        & Kp     	& 2.12           	& 0.351 		    	&  1.22"		&5				&3\\
\ \ SAO 146732	&       15:24        & H      	& 1.63            	& 0.296                     	&  1.23"		&5				&3\\
\ \ SAO 146732	&        15:27       & J     	& 1.25              	& 0.163                 	&  1.23"		&5				&3\\
28 July 2007	\\
\ \ Neptune  	& 11:28              & H      	& 1.63      		& 0.296                    	&  1.21"		&60				&5\\
\ \ Neptune	& 11:37              & H      	& 1.63        	& 0.296                       &  1.21"		&60				&5\\
\ \ Neptune	& 11:45              & H      	& 1.63        	& 0.296                       &  1.22"		&60				&5\\
\ \ SAO 146732	&       15:20        & H      	& 1.63            	& 0.296                       &  1.24"		&5				&3\\

\hline
\end{tabular}
}
\end{minipage}
\end{center}
\end{table*}

We obtained high spatial resolution NIR images of Neptune's south pole in July 2007, revealing temporal evolution of discrete circumpolar clouds. Here we describe these features, model their properties, and discuss the implications for the dynamics of the Neptunian south polar region.

\section{Observations and Data Processing}

We observed Neptune from the 10-meter W.M. Keck II telescope on Mauna Kea, Hawaii, on 26-28 July (UT) 2007, as part of a long-term program studying the planet's atmosphere in the NIR. J-, H-, and Kp-band images were taken using the narrow camera of the NIRC2 instrument, coupled to the adaptive optics (AO) system (Table 1). The 1024x1024 array has a pixel size of $9.963 \pm 0.011$ mas/pixel in this mode \citep{pravdo06}, which at the time of observations corresponded to a physical size of $\sim$210 km at disk center. 

The data were flat fielded and sky-subtracted, and bad pixels were replaced with the median of the 8 surrounding pixels. All images were corrected for the geometric distortion of the array using the `dewarp' routines provided by P. Brian Cameron\footnote{\url{http://www2.keck.hawaii.edu/inst/nirc2/post_observing/dewarp/nirc2dewarp.pro}}, who estimates residual errors at $\ltsimeq$ 0.1 pixels. We measured a full width at half maximum (FWHM) of $0.039\pm 0.005''$ for a stellar point source (SAO 146732) on the days of observation, which is consistent with the diffraction limit of the telescope at 2 $\mu$m, and corresponds to an effective resolution of $\sim$800 km at the center of the disk. 

The images were photometrically calibrated using the star HD 22686, then converted from units of observed flux density to units of $I/F$, which is defined as  \citep{hammel89}:
\[
\frac{I}{F} =\frac{r^2}{\Omega}\frac{F_N}{F_\odot}
\]
where $r$ is Neptune's heliocentric distance, $\pi F_\odot$ is the sun's flux density at Earth's orbit \citep{colina96}, $F_N$ is Neptune's observed flux density, and $\Omega$ is the solid angle subtended by a pixel on the detector. By this definition,  $I/F=1$ for uniformly diffuse scattering from a Lambert surface when viewed at normal incidence.

\begin{figure*}[htb!]
\begin{center}$
\begin{array}{ccccccccc}
\includegraphics[width=0.3\textwidth]{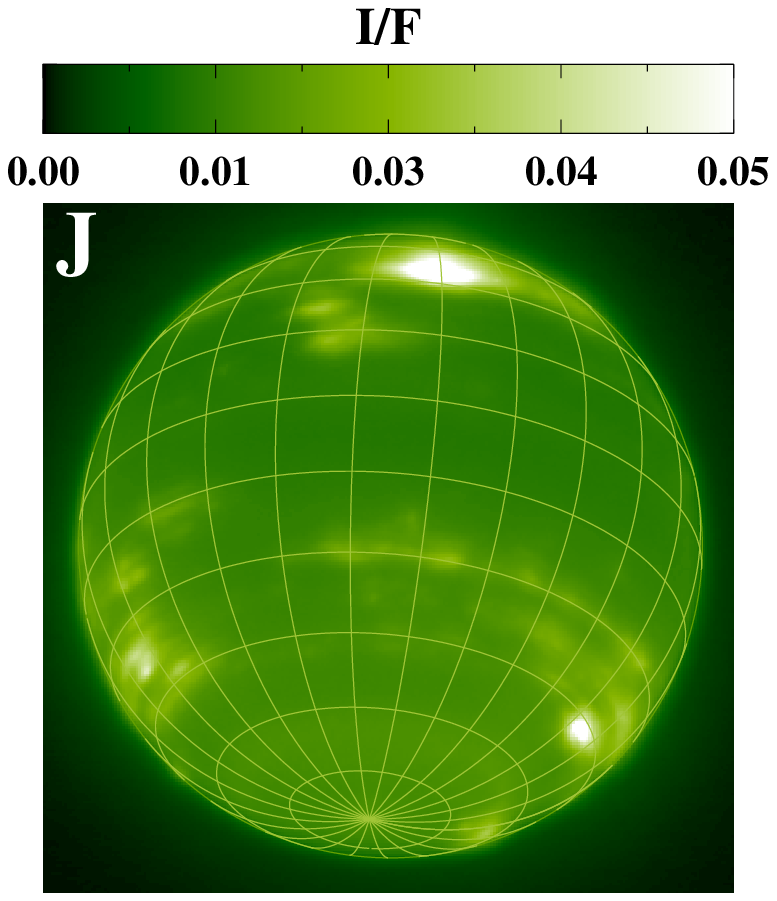}
\includegraphics[width=0.3\textwidth]{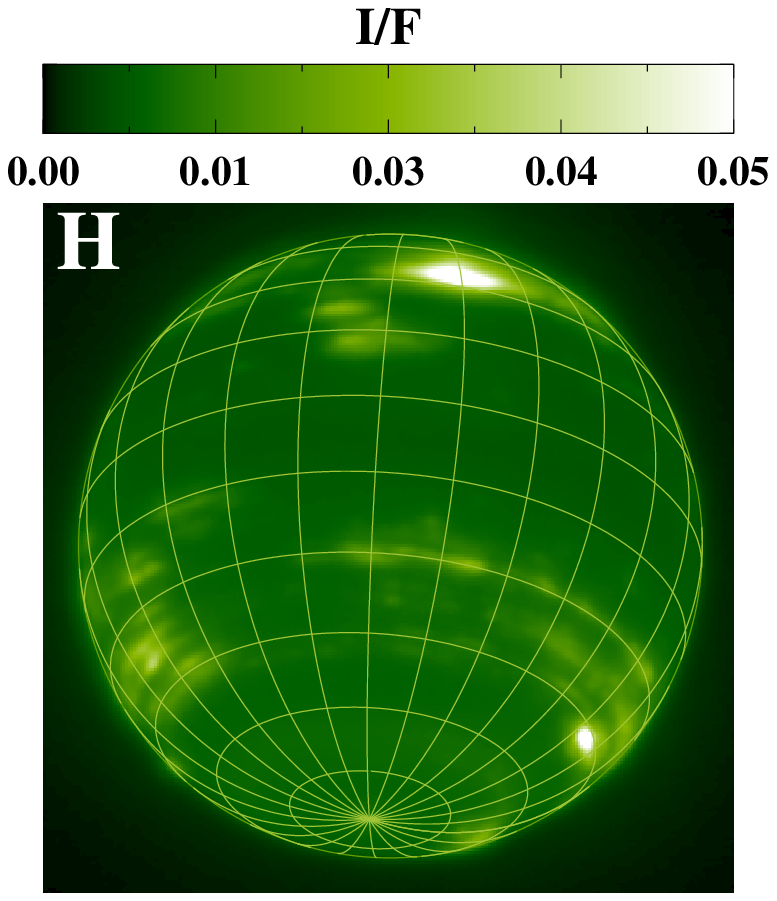}
\includegraphics[width=0.3\textwidth]{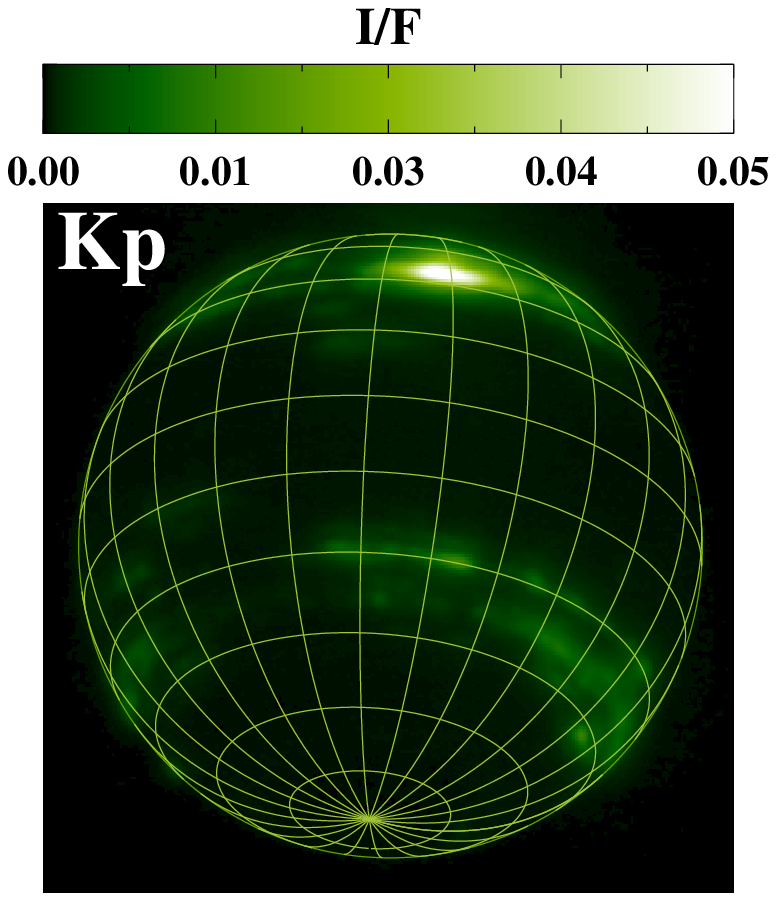}\\
\includegraphics[width=0.3\textwidth]{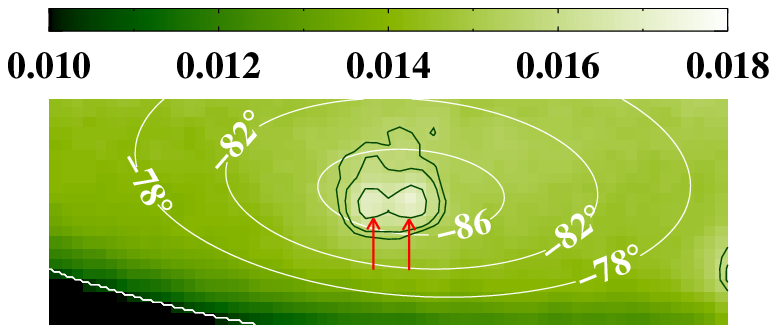}
\includegraphics[width=0.3\textwidth]{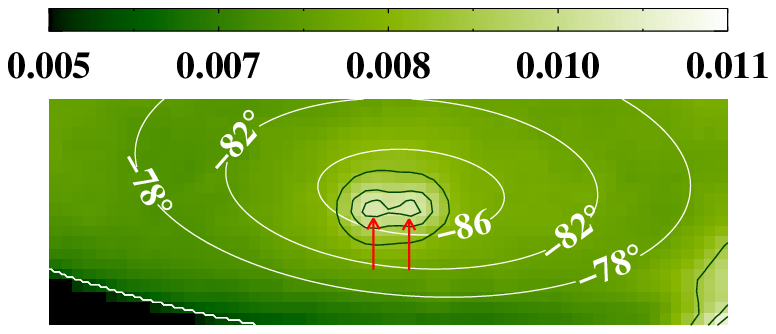}
\includegraphics[width=0.3\textwidth]{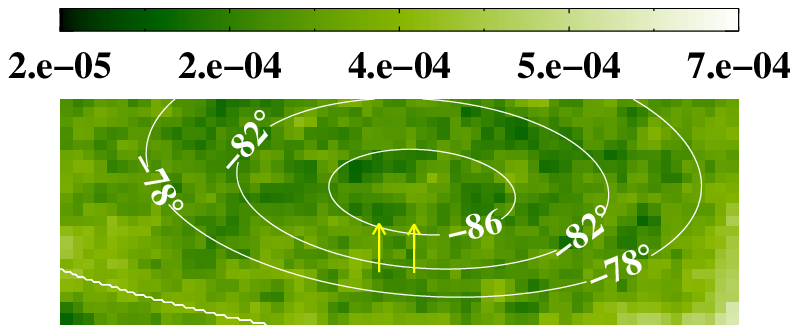}\\
\includegraphics[width=0.3\textwidth]{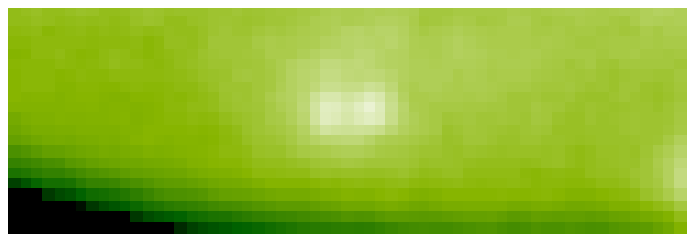}
\includegraphics[width=0.3\textwidth]{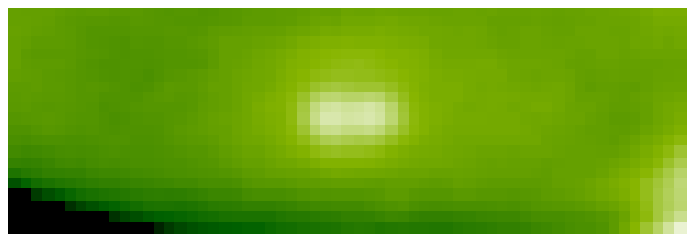}
\includegraphics[width=0.3\textwidth]{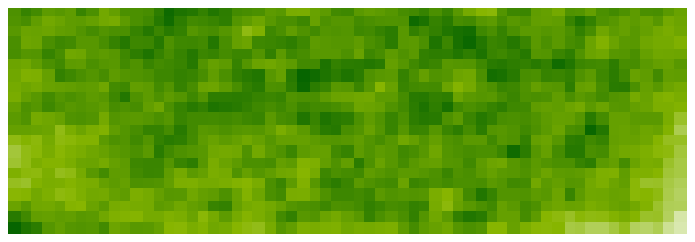}
\end{array}$
\end{center}
\caption{\label{jhk} J-, H- and Kp-band images from 26 July 2007 in units of I/F. The first three images in each filter were averaged together to produce the figure. The top panels all have same color bar. In the top images, the pole is bright due to the convergence of longitude lines; therefore we zoom in on the pole in the bottom panels.  The top row highlights the features by superposing contours at 10\% 20\% and 26\% above the H-band background; and 3\%, 5\% and 10\% above the J-band background level. Latitude circles at $4^{\circ}$ spacings are also shown. The arrows indicate the positions of the two cloud  features; in Kp band the clouds are not above the noise; therefore no intensity contours are plotted.\scriptsize }
\end{figure*}

\begin{figure}[htb!]
\begin{center}$
\begin{array}{cc}
\includegraphics[width=0.3\textwidth]{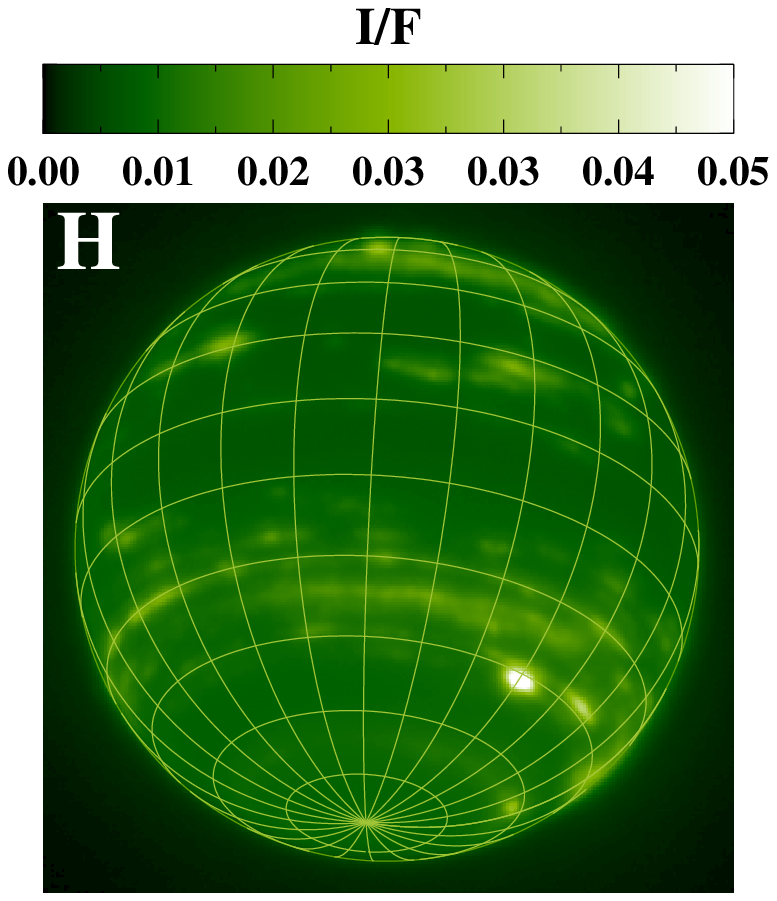}\\
\includegraphics[width=0.3\textwidth]{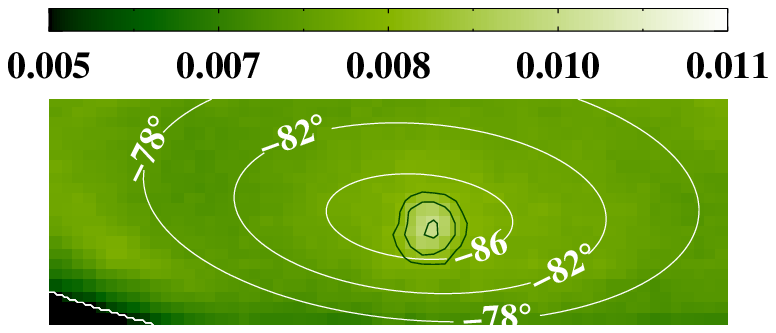}\\
\includegraphics[width=0.3\textwidth]{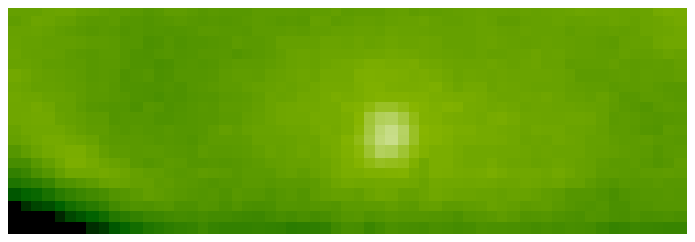}\\
\end{array}$
\end{center}
\caption{\label{nd} Average of first three H-band images from 28 July 2007, in units of I/F, with latitude and longitude lines plotted. The middle image highlights the single cloud feature with contours of 5\% 10\% and 20\% above the background level. \scriptsize}
\end{figure}

The 26 July data (Fig. \ref{jhk}) show two distinct features near Neptune's south pole. Other unresolved sources, such as moons and other clouds, do not appear double, so it seems unlikely this is an artifact. The 28 July data (Fig. \ref{nd}), at comparable resolution, show only a single unresolved polar cloud feature. To determine the total I/F of the unresolved features above the background, we model them as one (for the single cloud on 28 July) or two (for the pair of clouds on 26 July) 2D Gaussians. We use the best-fit parameters from the least-squares fitting routine MPFIT\footnote{Markwardt, C.B. 2008 in proc. Astronomical Data analysis Software and Systems XVIII} to find the I/F of each cloud feature (Table 2). From experience we estimate the error in the photometry to conservatively be 20\%; the errors in the cloud fits are between 4\% and 12\%.  In Kp band, we do not see the polar cloud features above the noise (Fig. \ref{jhk}). To set an upper limit for the I/F of the clouds in Kp band,  we assume the clouds will be detectible when the mean value of the signal plus noise is equal to one standard deviation above the noise mean. We confirmed this upper limit by convolving the maximum Kp I/F with a Gaussian having the FWHM of the point spread function (PSF) and adding the observed level of noise.

\section{Navigation and Cloud Locations}
\begin{figure}[htb!]
\begin{center}
\includegraphics[width=.4\textwidth]{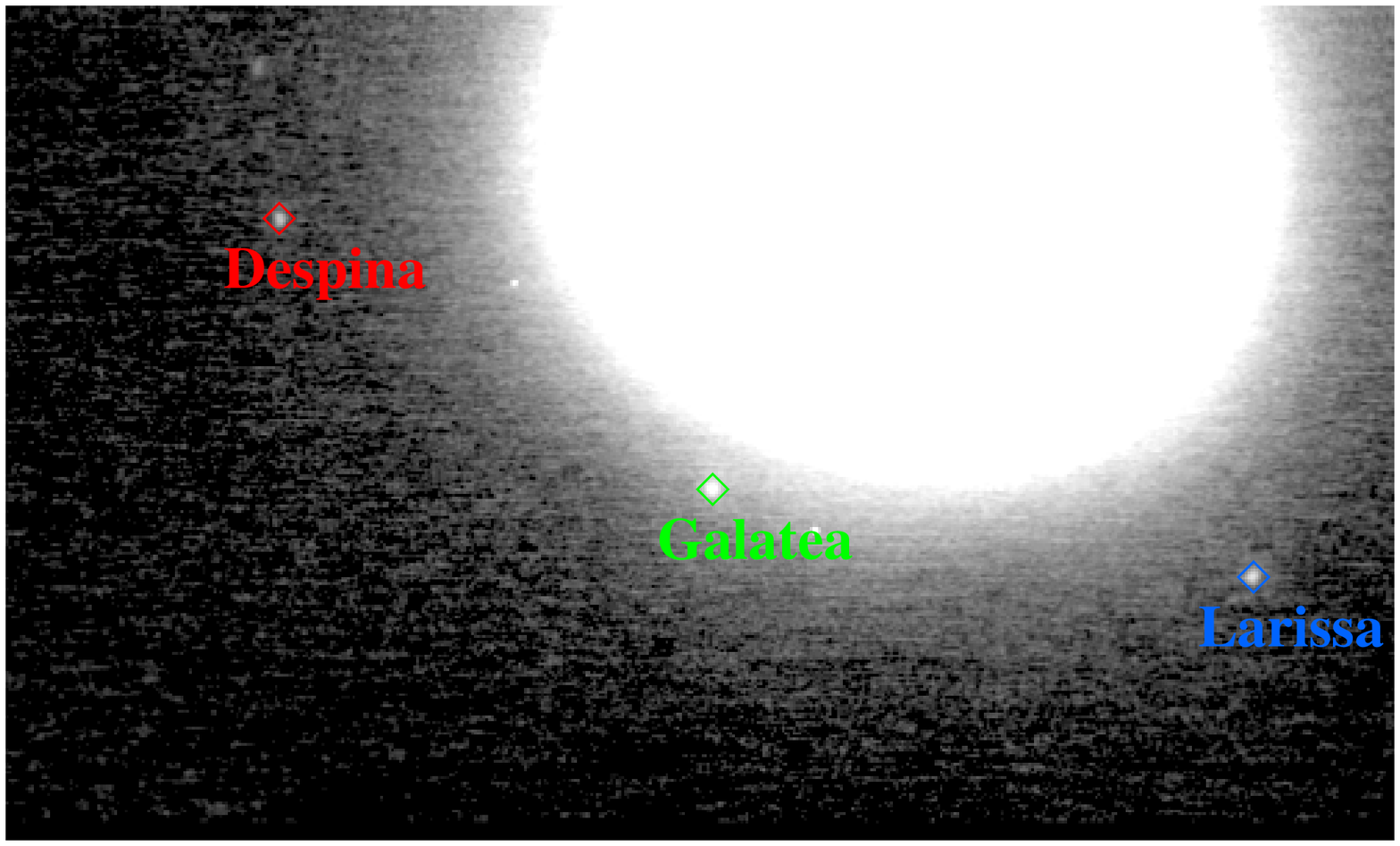}\\
\vspace{\baselineskip}
\includegraphics[width=.45\textwidth]{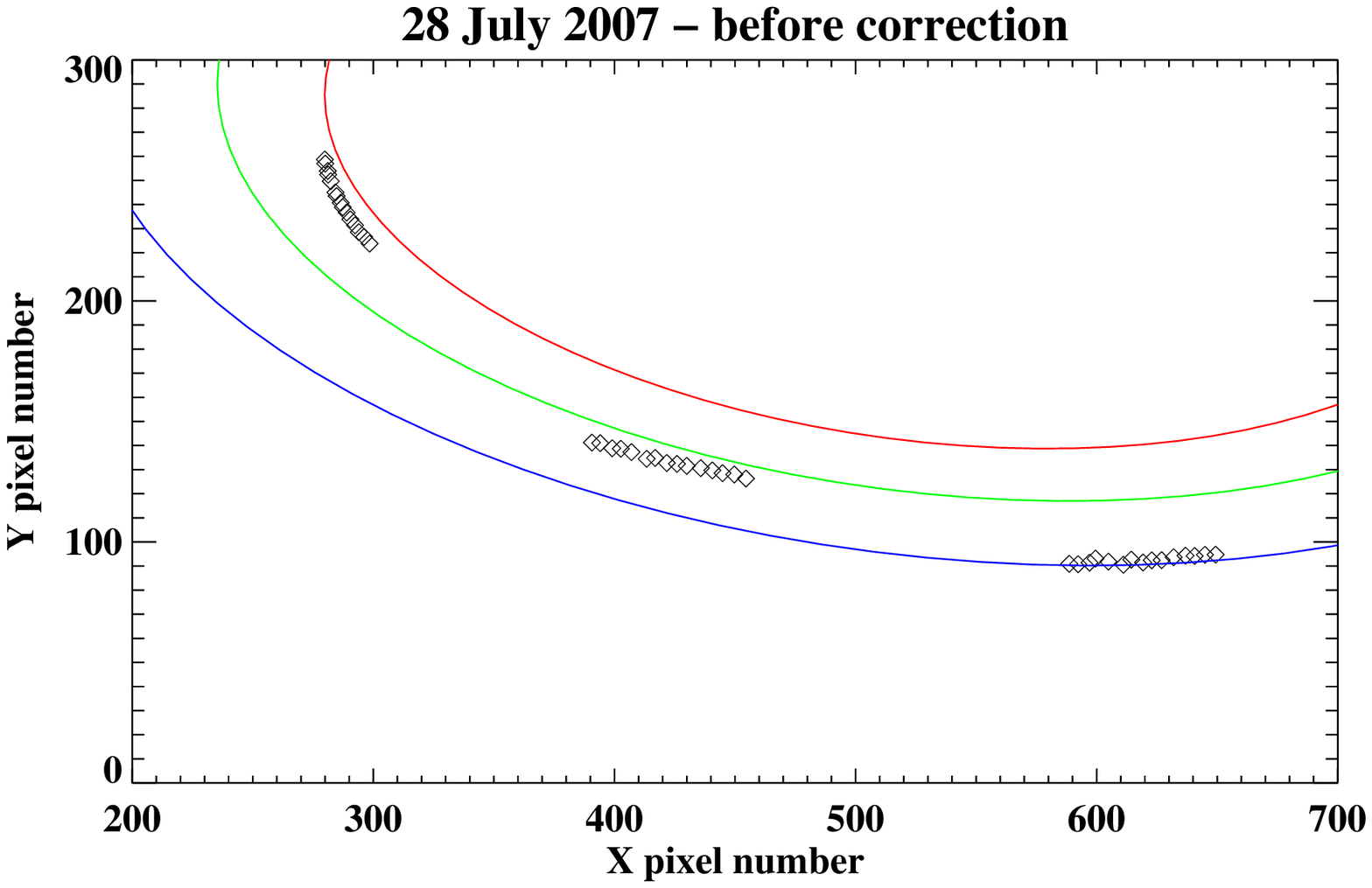}\\
\includegraphics[width=.45\textwidth]{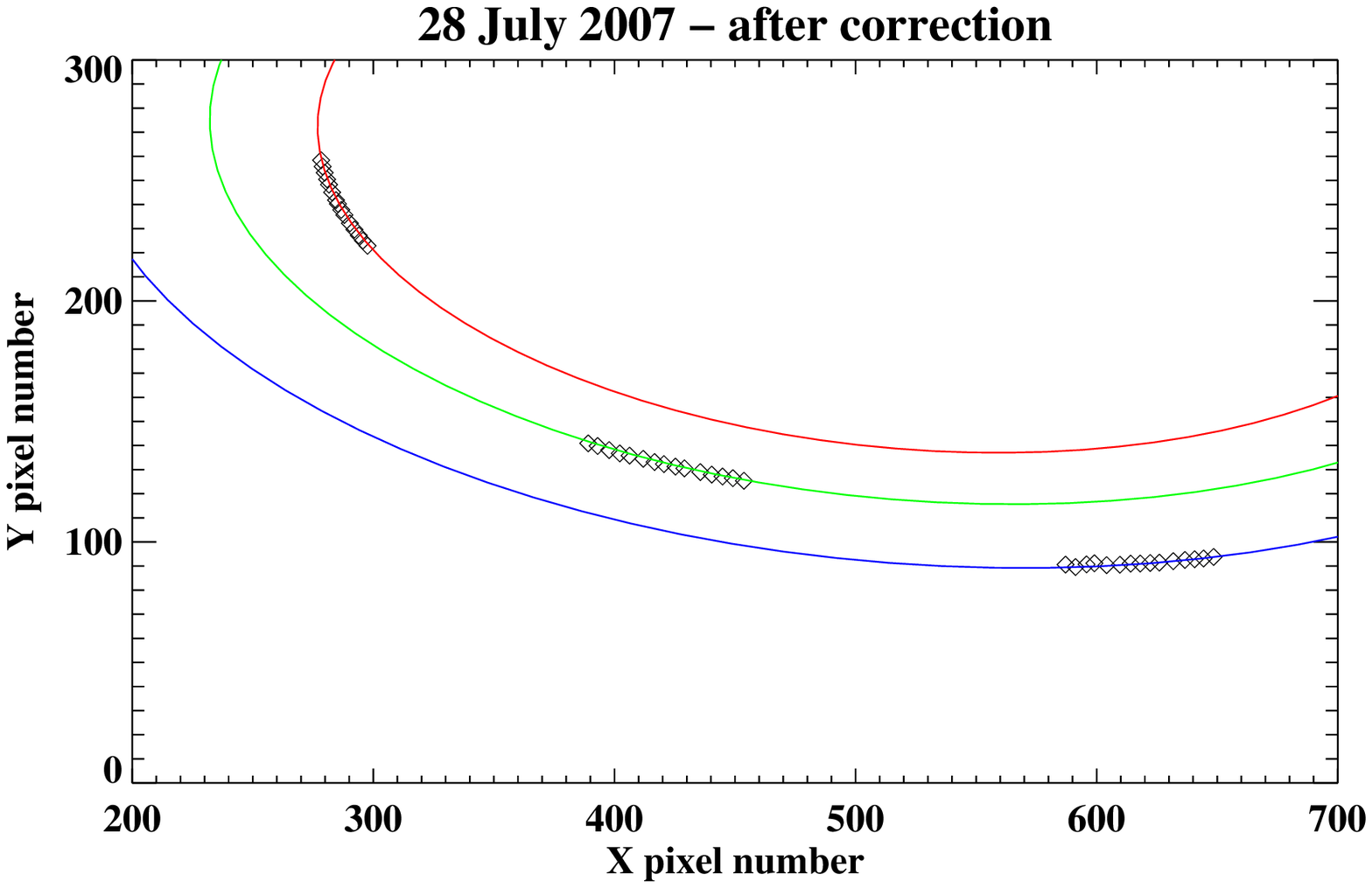}
\end{center}
\caption{\label{moonfit} Fitting disk center using Neptune's moons. The positions of Galatea (green), Larissa (blue) and Despina (red) were found in each image and aligned with the moon orbits to establish the planetary coordinates. The top panel shows the moons in the first H-band image on 28 July. The middle plot shows the orbits relative to the positions of the moons in the images when centered using the planet's limb. The bottom plot shows the moon positions after fitting them to their orbits. \scriptsize}
\end{figure}

Neptune's bright and dynamic activity limits the accuracy of image centering when using the planet's limb as a reference, since the PSFs of bright features near the limb often extend off the edge of the disk. To perform the centering more precisely, we found the positions of three of Neptune's moons in each H-band image, and shifted the images in $x$ and $y$ so that the positions of the moons were best fit to lie on their orbits, as derived from the Planetary Rings Node ephemeris data\footnote{\url{http://pds-rings.seti.org/tools/ephem2_nep.html}} (Fig. \ref{moonfit}).  While other methods of aligning images do not necessarily give you the location of disk center, this technique allows us to improve both the relative alignment of the images as well as the determination of disk center coordinates.  The centering error in our least-squares fit is on average 0.4 pixels per image, which agrees well with the image-to-image scatter of the positions of individual cloud features. This is an improvement over our centering by limb detection, which we estimate to result in 1--2 pixel errors in the determination of disk center. In J- and K- band we do not have a sufficient number of images to implement our moon centering technique; therefore these images  are aligned to the H-band images by eye, to sub-pixel accuracy. 

  \begin {table*}
 \begin{center}
 \caption{Spot locations and flux densities in units of I/F, as well as the I/F of the region surrounding the clouds. For the clouds, the maximum Kp/J and Kp/H ratios allowed by the data are also given. \scriptsize }
 \vspace{\baselineskip}
 \footnotesize{
\begin{tabular}{l l l l l l l}
\hline
					&					&		I/F	&		& 	& 	&\\
\hline
\ \ \ latitude 				& longitude			& J 		& H			 & Kp	&	max(Kp/J)	& 	max(Kp/H) \\
\hline
26 July 2007\\
\ \ $-89.08 \pm 0.06 ^{\circ}$	& $293 \pm 3 ^{\circ}$	 & $0.039\pm 0.009$     &$0.036\pm 0.008$&$<0 .004$  &$0.11\pm 0.02$	&$0.12 \pm 0.02$ \\
\ \ $-87.84 \pm 0.06 ^{\circ}$	& $238 \pm 2 ^{\circ}$	 & $0.038\pm 0.008$     &$0.033\pm 0.007$& $<0. 004$   &$0.12\pm 0.02$	&$0.13 \pm 0.03$ \\
\ \ background				&					&  $0.015\pm 0.003$ &$0.008\pm 0.002 $&$0.00023 \pm 0.00006   $    \\
28 July 2007\\
\ \ $-88.47\pm 0.03 ^{\circ}$& $86 \pm 2 ^{\circ}$	 & $0.017\pm 0.009$     &$0.042\pm 0.009$&$< 0.005$   &$0.3\pm 0.1$	&$0.12 \pm 0.01$ \\
\ \ background				&					&  $0.014\pm 0.003$ & $0.008\pm 0.002$&$0.00026 \pm 0.00009   $    \\
\hline
\end{tabular}
}
\end{center}
\end{table*}

To find the positions of the clouds near the pole, we use the moon-centered H-band images and the solutions from the 2D Gaussian fits (Section 2). We determine the weighted mean location of each cloud in image coordinates, which we then transform into planetocentric latitude and longitude, using the JPL Horizons ephemeris information\footnote{\url{http://ssd.jpl.nasa.gov/horizons}} for sub-observer latitude and longitude at the time of the observations. The errors in the latitude and longitude of the clouds are estimated using a Monte Carlo method, whereby for each parameter involved in calculating the latitude and longitude, a random distribution of values consistent with its probability distribution is generated. These parameters include the array rotation angle \citep{pravdo06}, image center coordinates, and cloud center coordinates. For each set of generated parameter values, the latitude and longitude are then calculated; errors are determined as the standard deviations of these simulated data sets. 

We find that on 26 July the two cloud features are near, but not directly at the pole as determined by moon orbits: the spot nearest the pole is located at $-89.07 \pm 0.06 ^{\circ}$ latitude and $293 \pm 3 ^{\circ}$ longitude. The second cloud resides  approximately 500 km further from the pole, at $-87.84 \pm 0.06 ^{\circ}$ latitude, $238 \pm 2 ^{\circ}$ longitude. The single spot on 28 July is likewise near, but not at, the south pole, at $-88.47 \pm 0.03 ^{\circ}$ latitude,  $86 \pm 2^{\circ}$ longitude (Table 2). 
 
\section{Radiative Transfer Modeling of Features}

\begin{figure}[htb!]
\centerline{\includegraphics[width=0.5\textwidth]{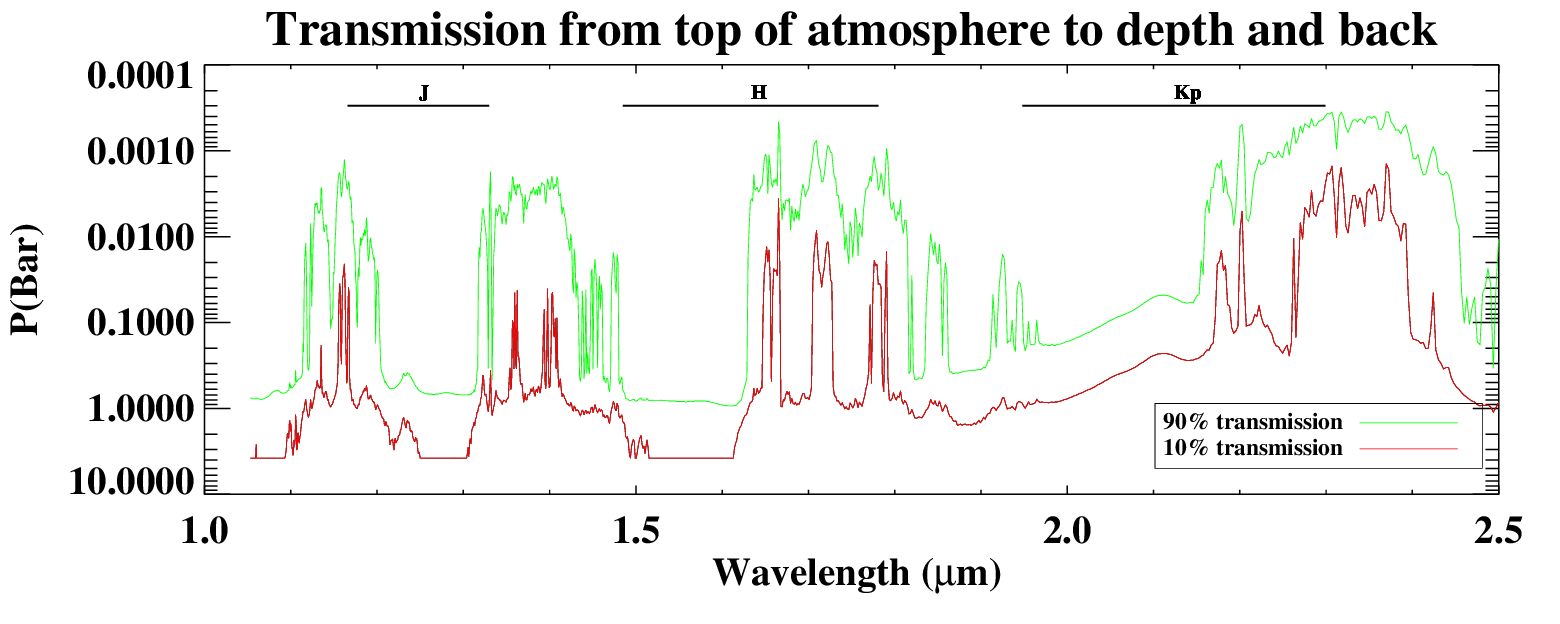}}
\caption{\label{trans}  Two-way transmission in Neptune's atmosphere as predicted by our RT model. The curves show the pressure levels at which 10\% and 90\% of the light is transmitted to the top of the atmosphere and back. \scriptsize}
\end{figure}

\begin{figure}[htb!]
\centerline{\includegraphics[width= 0.5\textwidth]{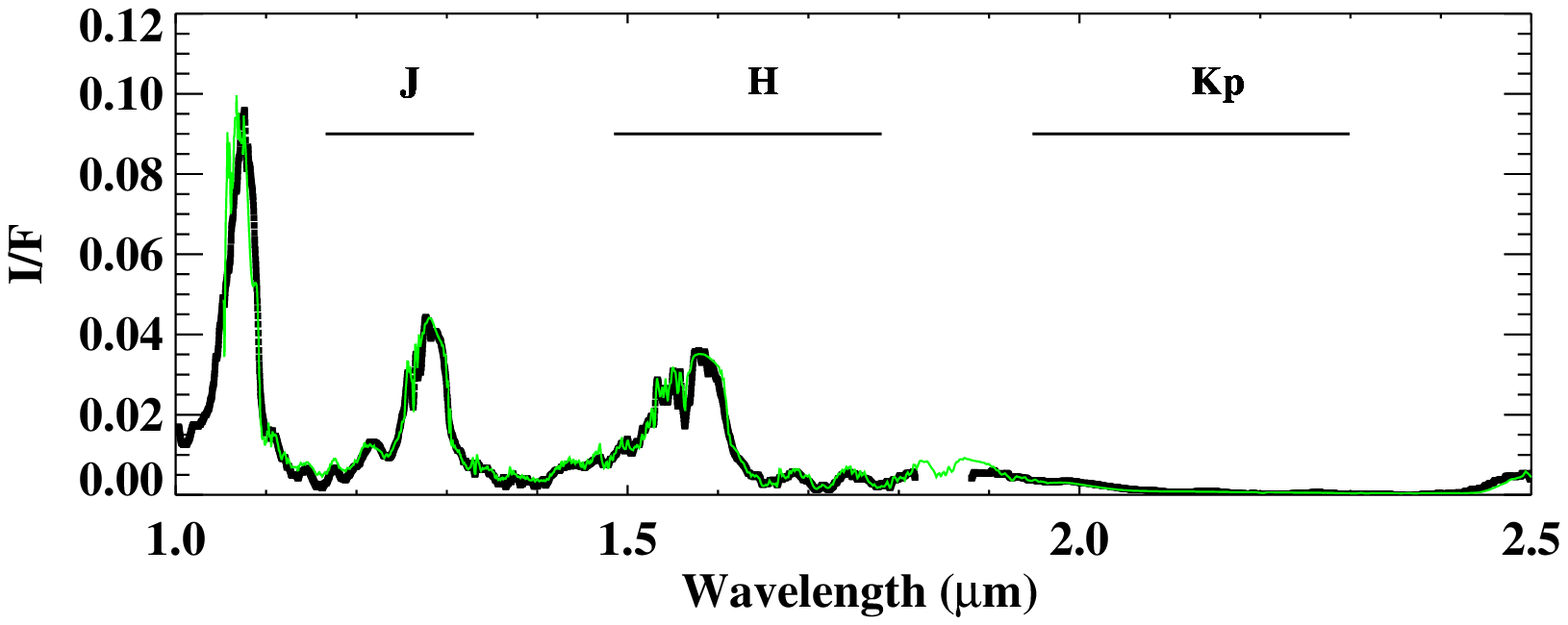}}
\caption{\label{model}  Comparison of radiative transfer model (green) and long-slit IRTF SpeX spectrum (black).\scriptsize }
\end{figure}

The J, H, and Kp filters probe various depths through the stratosphere and upper troposphere (Fig. \ref{trans}); therefore we can use the data to set limits on the heights of the south polar cloud features. To do this, we adapted the radiative transfer (RT) code for Titan from  \cite{mate07} for Neptune. We adopt the Neptune temperature profile derived by \cite{lindal92}  from occultation measurements, and incorporate the major opacity sources that dominate Neptune's 1- 5 $\mu$m spectrum:  we use the H$_2$ collision-induced absorption coefficients for hydrogen and helium \citep{borysow85,borysow88,borysow91,borysow92,borysow93}, assuming an equilibrium ortho/para ratio for H$_2$; and the k-distribution coefficients for H$_2$-broadened methane \citep{irwin06}. Following \cite{gibbard02}, our lower boundary is an optically thick cloud at 3.8 bar, presumably the H$_2$S cloudtop \citep{depater91}, with a Henyey-Greenstein asymmetry parameter of $-0.1$. Aerosols are treated as Mie scatterers with real indices of refraction of $1.43$ and imaginary indices of refraction $n_i=0$.  Within a given haze layer, aerosols are characterized by a single particle radius;  this radius, along with the altitude range and number density of aerosol particles, are free parameters, chosen to fit the data. The RT equations are then solved using a two-stream approximation \citep{toon89}.

To model the south polar cloud features, we first developed two cloud-free `background atmosphere' models. For the first model, we choose the simplest haze distribution consistent with the J-, H-, and K-band I/F values in the region surrounding the polar clouds (Table 3). For comparison, we also find a best model fit to a medium-resolution spectrum from the IRTF Spectral Library (Fig. \ref{model}), taken with the SpeX instrument on the NASA Infrared Telescope Facility (IRTF) on Mauna Kea \citep{rayner09}. These data are averaged over a significant fraction of Neptune's disk. Although this disk-integrated model has little contribution from the south pole, this second model, which has very different haze properties from the first (Table 3), provides a qualitative test of the sensitivity of our cloud modeling to the choice of model hazes. In addition, fitting to the SpeX data allows us to model the albedo of the deep `surface' cloud:  we find that given our choices of bottom cloud depth and scattering properties, the single-scattering albedo ($\omega$) of the cloud must decrease from $0.55$ at 1.1 $\mu$m to $0.20$ at 1.27 $\mu$m and $0.13$ at 1.56 $\mu$m. These values are reasonably consistent with \cite{roe01}.

\begin {table}[htb!]
\begin{center}
 \caption{Haze layers, chosen to fit the I/F in the three spectral windows in the the region around the polar clouds (top); and the IRTF SpeX data (bottom).\scriptsize }
 \vspace{\baselineskip}
 \footnotesize{
\begin{tabular}{l l l l }
\hline
\ \ P$_{min}$ (mbar)& P$_{max}$(mbar)& rad($\mu m$)  &n(cm$^{-3})$ \\
\hline
south polar region fit\\
\ \ 5		&10		&0.8			&0.002\\
\ \ 45		&50		&0.2			&100\\
\ \ 540	&1540	&0.5			&3\\
SpeX fit \\
\ \ 1.4	& 20	 	&0.2     		&12.5 \\
\ \ 340	&1540	&2.5		         &0.07\\
\hline
\end{tabular}
}
\end{center}
\end{table}

We then model each of the south polar clouds as a perfectly reflecting ($\omega=1$), optically thick cloud layer with a Mie asymmetry parameter of 0.5. We consider clouds at each of 100 altitudes between 0.4 mbar and 3.8 bar. For each case, we average the resulting 'cloudy' model over H, J, and Kp bands for comparison with our image data, and subtract the I/F of the `cloud-free' model to get the I/F contribution from the cloud.

\begin{figure} [htb!]
\includegraphics[width=0.45\textwidth]{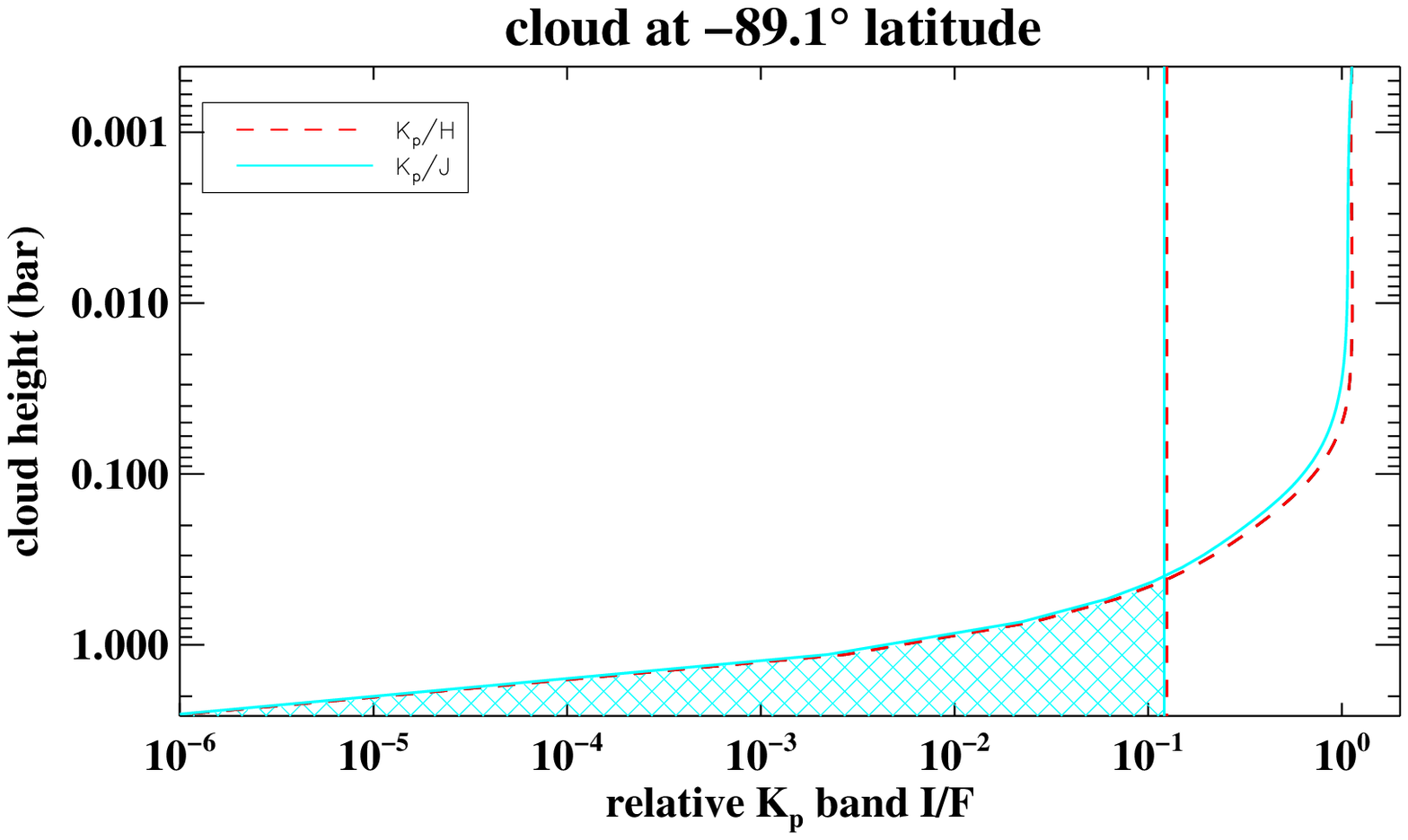}
\includegraphics[width=0.45\textwidth]{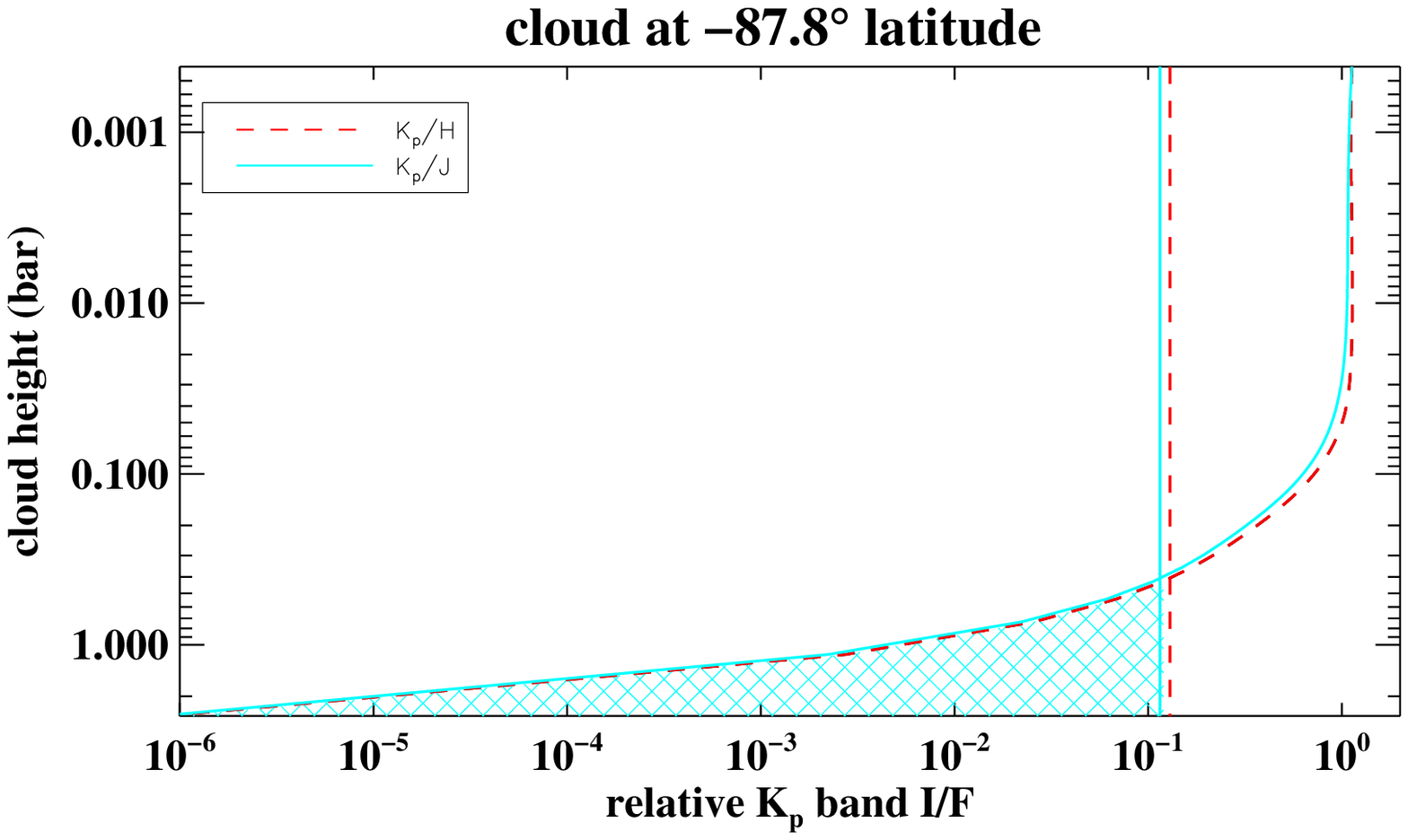}
\caption{\label{relif}  These plots show the model results for Kp/H intensity ratio (solid red) and Kp/J ratio (solid cyan) as a function of cloud altitude for each of the two cloud features from 26 July 2007. Dotted lines are at the maximum ratios allowed by the data, showing that the Kp/H ratio sets a stronger limit on the maximum cloud height than the Kp/J ratio. The allowed parameter space is cross-hatched.\scriptsize }
\end{figure}

For the 26 July data, we find the ratio of the intensities of the two clouds, which is affected by spot fitting errors but not calibration errors, to be $1.1\pm 0.1$ in H band and $1.0\pm 0.1$ in J band. Therefore, our results are consistent with the clouds being of the same size and at the same altitude. However, since the clouds are unresolved, we do not know the pixel filling fraction for each of the clouds. We can set an upper limit on the clouds' altitudes by comparing the maximum model-derived Kp-to-H and Kp-to-J I/F ratios to the observed values: the higher a cloud is in the atmosphere, the greater the expected Kp-band intensity. The upper limit Kp-to-H and Kp-to-J intensity ratios for each of the clouds, given the uncertainties in H and J, are presented in Table 2. Figure \ref{relif} shows the results of the modeled Kp/H and Kp/J intensities, compared to the maximum of these ratios from the 26 July data.

We find that the upper limit to the altitude of both clouds on 26 July  is 0.4 bar. This also provides an estimate on the minimum size of the clouds: if either of the clouds is at 0.4 bar, it must fill at least 10\% of a NIRC2 narrow camera pixel to be consistent with the observed H and J-band intensities. A deeper cloud would necessarily be larger to produce the same values of I/F. For the single cloud on 28 July, we find that its altitude also has an upper limit of 0.4 bar. 

The models used in Fig. \ref{relif} use the haze distribution that best matches the I/F of the background atmosphere near the cloud features. However, we expect our choice of haze parameters to have little effect on our results. This is because we are interested primarily in the I/F increase due to adding a cloud into the background `cloud-free' atmosphere. When we ran our second set of models using the simpler haze distribution used to fit the IRTF SpeX data (Table 2), we found that changing the haze parameters affected the `cloud-free' and `cloudy' spectra in a similar way, so that to within 3\%, the calculated change in I/F from a given cloud was independent of the haze parameters.

\section{Discussion}

A small bright feature at Neptune's south pole has been consistently observed since the Voyager era, implying a stability that contrasts the short ($\sim$hours) lifetime of most small cloud features at lower latitudes \citep{limaye91}. Our derived upper limit of the altitude of the circumpolar features observed here is consistent with clouds formed by the upwelling and condensation of methane gas, suggesting this may be a region of continuous cloud formation by convection.

The persistence of south polar cloud activity suggests that there is an organized circulation pattern at Neptune's south pole. Such  a pattern has been inferred before: \cite{hammel07b} show that there is a mid-IR temperature enhancement at Neptune's south pole, and \cite{martin08} find that the pole is warmer than its surroundings at microwave  wavelengths as well. The latter authors 
suggest that a global dynamical pattern, in which air near the south pole is subsiding in both the stratosphere and troposphere, can explain the observed high temperatures at both the mid-IR and radio wavelengths. Subsiding motions will adiabatically heat the atmosphere, visible as elevated temperatures in the mid-IR. The subsiding air is likely ÓdryÓ, as condensable gases will condense out during the ascending branch of the dynamical pattern. Such dry conditions enables one to probe much deeper warmer levels at radio wavelengths, leading to temperature enhancement in microwave observations. 

\begin{figure} [htb!]
\centerline{\includegraphics[width=0.45\textwidth]{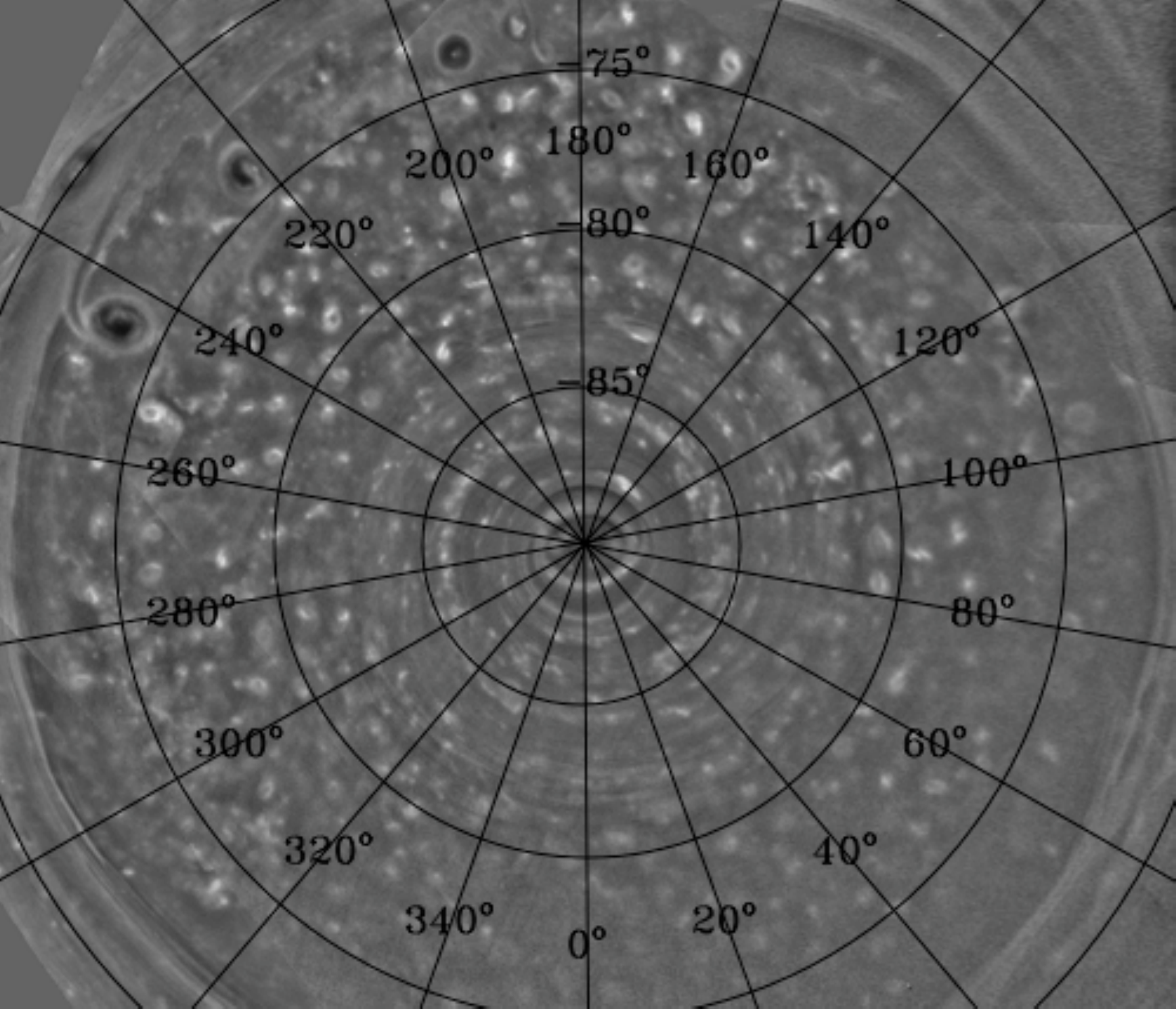}}
\caption{\label{sat} Map of Saturn's SPV [adapted from \cite{dyudina09}] illustrating the small, bright clouds encircling the storm's `eye', many of which have the same general appearance between 2004 and 2006 observations. \cite{dyudina08} conclude these clouds are located in Saturn's lower troposphere.\scriptsize }
\end{figure}

On Saturn, a south polar circulation cell has been observed, in the form of subsidence in a polar vortex surrounded by a region of upwelling \citep{fletcher08}. The structure of Saturn's south polar vortex (SPV) \citep{dyudina09} possesses similarities with terrestrial hurricanes, such as a well-formed central eye, concentric eyewalls, and a surrounding ring of strong convection (Fig. \ref{sat}). The region corresponding to the eye of the SPV is a hot, nearly circular region within a diameter of 4200 km around the south pole. The eye is mostly clear of clouds above $\sim$1 bar \citep{dyudina08}, and depleted in phosphine gas \citep{fletcher08}, suggesting that this is a region of subsiding air.  Surrounding the eye is a ring of discrete, bright clouds in the lower troposphere, which may be analogous to the heavily precipitating clouds encircling the eye of terrestrial hurricanes \citep{dyudina09}. The similarity between the temperature enhancement at Neptune's south pole to the enhancement at Saturn's south pole \citep[and references therein]{hammel07b} first led to the suggestion that Neptune, like Saturn, harbors a long-lived hot south polar vortex  \citep{orton07}.   Our finding that Neptune's south polar cloud feature(s) may be indicative of rigorous convection near, but not at, the pole, supports the analogy between Neptune and Saturn's south polar environments.

The unexpected discovery of a vortex at Saturn's north (winter) pole \citep{fletcher08} implies that polar vortices can exist despite considerable variations in seasonal insolation, and may be general
features of giant planet atmospheres. However, possible analogues to Saturn's vortices have yet to be directly observed, as they await the measurement of polar wind speeds by high-inclination space missions. Further details about the morphology of clouds near Saturn's north pole may help us understand the types of cloud morphologies we might expect around polar vortices, and provide insight into how to interpret the behavior of Neptune's polar cloud features in the context of a possible vortex.

\section{Summary}
We have observed a transient double cloud feature near Neptune's south pole on 26 July 2007. The locations of the two features, as determined by the orbits of three of Neptune's moons, suggest they are both circumpolar clouds separated in latitude by $\sim1.2^{\circ}$. The single circumpolar feature seen on 28 July 2007 is at an intermediate latitude. Radiative transfer modeling  indicates that the features on both days are at depths of greater than 0.4 bar, which is consistent with the formation of methane condensation clouds. The morphological change we observe on 26 July, combined with the location of the single cloud on 28 July, may indicate that the bright south polar spot seen since Voyager is not a single stable feature, but rather the signature of persistent cloud activity related to strong convection, perhaps in the region surrounding a polar vortex. Continued high-resolution observations of Neptune's south pole will  provide more insight into the dynamics of this region. Such observations should be performed over several hours to capture the evolution of the features. We also await more high-inclination observations of planetary atmospheres to provide further context for understanding giant planet polar environments.

\begin{acknowledgements}
The  data presented in this work were obtained with the W.M. Keck Observatory, which is operated by the California Institute of Technology, the University of California, and the National Aeronautics and Space Administration. The Observatory was made possible by the generous financial support of the W.M. Keck Foundation. This work was supported by the National Science Foundation Science and Technology Center for Adaptive Optics, managed by the University of California at Santa Cruz under cooperative agreement AST 98-76783. Further support was provided by NSF grant AST-0908575. HBH acknowledges support for this work from NASA grants NNX06AD12G and NNA07CN65A. The authors extend special  thanks to those of Hawaiian ancestry on whose sacred mountain we are privileged to be guests. Without their generous hospitality, none of the observations presented would have been possible.
\end{acknowledgements}

\bibliographystyle{apalike}
\bibliography{refs}

\end{document}